\documentclass[12pt]{article}
\input epsf.sty
\def\be{\begin{equation}}
\def\ee{\end{equation}}
\def\ba{\begin{align}}
\def\ea{\end{align}}

\def\p{\partial}

\def\ops[#1]{\p_{#1} e^{-2\phi}}

\def\eq[#1]{equation (\ref {eq:#1})}
\def\Eq[#1]{Equation (\ref {eq:#1})}
\def\e[#1]{\ref {eq:#1}}
\def\at[#1]{| _{#1}}

\let\oldpercent\%\renewcommand{\%}{\scalebox{0.85}{\oldpercent}}

\begin{document}

\baselineskip=18pt

\begin{center}
{\Large \bf{On the cigar CFT and Schwarzschild horizons}}

\vspace{10mm}

\textit{ Amit Giveon}
\break

Racah Institute of Physics, The Hebrew University \\
Jerusalem, 91904, Israel

\end{center}


\vspace{10mm}

\begin{abstract}

Aspects of shock waves and instantly-created folded strings operators in the supersymmetric $SL(2)_k/U(1)$ CFT,
and their relevance to the near-horizon physics of Schwarzschild black holes in perurbative superstring theory, are presented.

\end{abstract}
\vspace{10mm}

In this short note, we provide more support to the claim \cite{Giveon:2021gsc}
that the properties of the {\it exact} cigar CFT hints towards the nature
of explicit microstates at the Schwarzschild black-holes horizons in perturbative string theory.

Recently, two interesting vertex operators showed up in the study of the {\it bosonic} cigar CFT.
One, denoted $F$ in~\cite{Giveon:2019gfk,Giveon:2019twx},
was interpreted as corresponding to the condensate of the instantly-created folded strings,
presented in~\cite{Itzhaki:2018glf,Attali:2018goq},
near the horizon of the black hole,
and the other, presented in~\cite{Chen:2021emg}, amounts to a shock wave that travels along the horizon
and is thus related to the chaos (or Lyapunov) exponent.

Here, we shall consider these operators, and the relation between them, in the {\it supersymmetric} cigar CFT, $SL(2)_k/U(1)$,
and what it is likely to apply to the near-horizon physics of $D$-dimensional black holes in the perturbative {\it superstring} theory.

While there isn't a bottom line, a speculative claim will await to the end.

The technical bottom lines are:
\begin{enumerate}
\item
Both the shock wave operator and the instantly-created folded strings operator, $F$,~\footnote{We shall denote by $F$ also the
supersymmetric version of the one in~\cite{Giveon:2019gfk,Giveon:2019twx}.}
are `fusions' of the $N=2$ Liouville supervertex operator $W$ with another $(1/2,1/2)$ superoperator $V$,
\be\label{owv}
O(w,\bar w)\sim\int d^2w_1d^2\theta_1 d^2\theta_2 W(w_1,\bar w_1;\theta_1,\bar\theta_1)V(w,\bar w;\theta_2,\bar\theta_2)
\ee
e.g. $O=F$ if $V=W^*$, while $O$ is the chaos exponent operator for a certain $V$ that we shall present below.
\item
When the $SL(2)$ level $k$ is a half, a quartet of things happen.
First, the shock wave operator coincides with the folded strings condensate operator $F$ when $k=1/2$.
Second, $k=1/2$ is  the level which amounts to that of the $2d$ fermionic string on the cigar SCFT,~\cite{Giveon:2003wn}, a.k.a.
\be\label{c}
c(SL(2)_k/U(1))=3+{6\over k}
\ee
is equal to $15$ when $k=1/2$.
Finally, when $k=1/2$, both the chaos exponent, $\lambda_L(k)$, in the shock wave operator
and the folded strings condensate value, $\lambda_F(k)$, {\it vanish}, a.k.a.
\be\label{lambdal}
\lambda_L(k=1/2)=\lambda_F(k=1/2)=0
\ee
\item
The exponential growth, $e^{\lambda_L(r_0)t}$ with time $t$,
of scattering amplitudes at the horizons of large $D$-dimensional Schwarzschild black holes in perturbative superstring theory,
a.k.a. with horisons size $r_0\gg$ string length scale $\ell_s$,
obtained using the methods of~\cite{Shenker:2014cwa},
is in coherence with the large $k$ behavior of the shock wave operator, characterized algebraically and obtained as in~\cite{Chen:2021emg}
in the cigar SCFT, a.k.a. with the exponential growth of amplitudes, $e^{\lambda_L(k)t}$,
at the horizon of the supersymmetric $SL(2)_k/U(1)$ black hole.
\end{enumerate}

To show the above, we first recall that the $N=2$ Liouville operator in a supersymmetric linear-dilaton cylinder,
$R_\phi\times S^1_x$ with dilaton $=-{Q\over 2}\phi$, is
\be\label{intw}
\int d^2\theta W=\psi\bar\psi w+c.c.
\ee
where
\be\label{w}
W=e^{-{1\over Q}Z}
\ee
is the $N=2$ Liouville superpotential written in terms of the chiral superfield
\be\label{z}
Z=z+\theta\psi+\bar\theta\bar\psi+\dots
\ee
with
\be\label{zpsi}
z\equiv\phi+ix~,\qquad
\psi\equiv\psi^\phi+i\psi^x
\ee
and~\footnote{Here we consider a momentum mode, as opposed to the winding {\it condensate} in the cigar.}~\footnote{In
the T-dual of the cigar CFT, such an operator amounts to a momentum condensate in the trumpet.
Recall,~\cite{Giveon:1991sy} (see also section 4.5 of~\cite{Giveon:1994fu}),
that upon continuation to Lorentzian spacetime, T-duality exchanges the horizon with the singularity.
In what follows, most of the heros in the intermediate manipulations will have neither well defined windings and/or momenta
on both a cigar and/or a trumpet. Nevertheless, upon continuation to Lorentzian spacetime of the instantly-created folded strings
and the shock waves operators, obtained in the bottom lines, not only both are likely to play important roles
in the near-horizon physics of black holes in string theory,
but the way they are fused below may also point, via the duality recalled above, towards revealing the physics near singularities.}
\be\label{smallw}
w\equiv e^{-{1\over Q}z}
\ee

In the supersymmetric $SL(2)_k/U(1)$ CFT, whose asymptotic regime is the linear-dilaton supercylinder above, with $x\simeq x+2\pi\sqrt{2k}$,
the slope $Q$ of the dilaton is related to the level $k$ by
\be\label{q}
Q\equiv\sqrt{2\over k}
\ee
where we set above and below
\be\label{alphap}
\alpha'\equiv\ell_s^2=2
\ee
when not presented explicitly.

Next, let
\be\label{v}
V\equiv e^{q(\phi+...)+ip(x+...)}
\ee
be a $(1/2,1/2)$ supervertex operator,~\footnote{The `...' in the exponent
stand for the higher components of the superfields whose lowest components are $\phi$ and $x$, respectively.} a.k.a.
\be\label{pq}
p^2=1+q(q+Q)
\ee
and define
\be\label{qpq}
Q\gamma\equiv p-q
\ee
One finds~\footnote{As in (3.22),(3.23) of~\cite{Brower:2006ea};
see the Pomeron operator in the appendix.} that
\be\label{wvo}
\int d^2w_1d^2\theta_1 d^2\theta_2 W(w_1,\bar w_1;\theta_1,\bar\theta_1)V(0;\theta_2\bar\theta_2)\sim 2\pi{\Gamma(1-\gamma)\over\Gamma(\gamma)}
\left(k\over 2\right)^\gamma O_\gamma(0)+\dots
\ee
where the $(1,1)$ operator $O_\gamma(w,\bar w)$ is
\be\label{o}
O_\gamma\equiv(\partial z\bar\partial z)^\gamma
e^{\left(p_\phi-{1\over Q}\right)\phi+i\left(p_\phi+Q\gamma-{1\over Q}\right)x}
\ee
and the `$\dots$' stand for fermionic and/or subleading terms.

Two special cases are the following.

The first is
\be\label{vwstar}
V=W^*=e^{-{1\over Q}Z^*}
\ee
a.k.a.
\be\label{pqwstar}
(q,p)=\left(-{1\over Q},{1\over Q}\right)\,\,\Leftrightarrow\,\,\gamma=k
\ee
and
\be\label{of}
O_\gamma=(\partial z\bar\partial z)^ke^{-{2\over  Q}\phi}
\ee
is $x$-independent; it is the leading asymptotic behavior of the folded strings operator $F$ in~\cite{Giveon:2019gfk,Giveon:2019twx}
in the asymptotic regime of the supersymmetric cigar CFT.~\footnote{In~\cite{Giveon:2019gfk,Giveon:2019twx},
$F$ was obtained in the bosonic case, hence, for the comparison of $F$ here to the one in~\cite{Giveon:2019gfk,Giveon:2019twx},
the level $k$ factors in some places above should be replaced by the bosonic level $k+2$. Moreover,
$F$ in~\cite{Giveon:2019gfk,Giveon:2019twx} was obtained as the fusion of winding and anti-winding operators $W^\pm$,
where $W^\pm$ are related to the bosonic sine-Liouville version of $W,W^*$ by $x=x_L+x_R\to\tilde x=x_L-x_R$, a.k.a. T-duality
(for a review, see~\cite{Giveon:1994fu}).}

The second case is
\be\label{vcm}
V=e^{Q{k-1\over 2}(\phi+...)+iQ\left({k-1\over 2}+\gamma\right)(x+...)}
\ee
a.k.a.
\be\label{gammacm}
\gamma={1\over 2}\left(1-k+\sqrt{k^2+2k-1}\right)
\ee
and
\be\label{ocm}
O_\gamma=(\partial z\bar\partial z)^\gamma e^{-{Q\over 2}\phi+iQ\left(\gamma-{1\over 2}\right)x}
\ee
is the behavior of the shock wave operator, constructed as in~\cite{Chen:2021emg},
in the asymptotic regime of the supersymmetric cigar,~\footnote{This can be shown
by extending the manipulations in section 2.5 of~\cite{Chen:2021emg}, section 5 of~\cite{Giveon:2016dxe} and section 3 of~\cite{Giveon:2019twx}
to the supersymmetric $SL(2)_k/U(1)$ CFT; we will not present the details here.}~\footnote{Extending
the manipulations in~\cite{Giveon:2016dxe} to the supersymmetric case, using e.g.~\cite{Giveon:2003wn},
the following can be shown. The T-duals of~(\ref{ocm}) and the top component of~(\ref{vcm}), a.k.a. $x\to\tilde x$ in both
(see~\cite{Giveon:1994fu}, for a review),
are related by a formal continuation of the supersymmetric version of GFZZ correspondence,~\cite{Giveon:2016dxe,Martinec:2020gkv},
to operators with the following $SL(2)_k/U(1)$ quantum numbers, $j,m={1\over 2}(n+k\omega),\bar m={1\over 2}(-n+k\omega)$.
The T-duals of~(\ref{ocm}) and the top component of~(\ref{vcm}) have $j=-{1\over 2}$ and ${k-1\over 2}$, respectively,
a.k.a. at the two unitarity bounds, and $m=\bar m={k\omega\over 2}=j+\gamma$, a.k.a non-integer windings, $\omega$, on the cigar.
Correspondingly, the shock wave operator $O_\gamma$ in~(\ref{ocm}), and its T-dual,
also have a non-integer left and right excitation number, $\gamma$.
We will not present the details of the GFZZ map between the T-duals of~(\ref{ocm}) and the top component of~(\ref{vcm}) here.
And, although we did not pin down the significance of this intriguing observation, it is tempting to point out again the following.
Recall,~\cite{Giveon:1991sy} (see also section~4.5 of~\cite{Giveon:1994fu}),
that upon continuation to Lorentzian spacetime, the T-duality used above exchanges the horizon with the singularity.
The observation in this footnote is thus an addition to other facts presented in this note,
which may have relevance also to the physics near singularities in string theory.}
whose $x$ momentum, $Q\left(\gamma-{1\over 2}\right)$,
amounts to an exponential growth $e^{\lambda_Lt}$ of amplitudes at the horizon of the supersymmetric $SL(2)_k/U(1)$ black hole,
upon the continuation $t\to ix$, with Lyapunov exponent
\be\label{llcm}
\lambda_L={2\pi\over\beta}b=Q\left(\gamma-{1\over 2}\right)
\ee
where we used the fact that, similar to~\cite{Chen:2021emg}, also in the supersymmetric case, the boost eigenvalue of the operator above,
in the Lorentzian theory, is
\be\label{b}
b=2\gamma-1
\ee
and that the Hawking inverse temperature is
\be\label{beta}
\beta=2\pi\sqrt{2k}
\ee

At large $k$,
\be\label{blargek}
b=1-{1\over k}+\dots
\ee
as in the bosonic case in~\cite{Chen:2021emg}.
In particular, $b\to 1$ when $k\to\infty$,
in which limit the shock wave operator $\to\partial z\bar\partial z$,
a.k.a. the Euclidean continuation ($t\to ix$) of a graviton shock wave in flat two-dimensional spacetime,
and the leading curvature correction to $b$ is $-1/k$.

On the other hand, as $k$ decreases, $b$ decreases,
and it approaches its smallest value, $b\to 0$, when $k\to 1/2$,
as follows from~(\ref{b}) with~(\ref{gammacm}).
Thus, at the value of $k$ for which the Lyapunov exponent~(\ref{llcm}) vanishes, the shock wave operator in~(\ref{ocm}) is
\be\label{ohalf}
O_{1/2}=(\partial z\bar\partial z)^{1/2} e^{-\phi}
\ee
which is precisely the instantly-created folded string operator in~(\ref{of}), when $k=1/2$, as advertised above.

String theory on the $SL(2)_{k=1/2}/U(1)$ black-hole sigma model is below the black hole/string transition point at $k=1$,~\cite{Giveon:2005mi}.
It was argued that the description of the worldsheet $SL(2)_k/U(1)$ SCFT in terms of the
geometry of a cigar is misleading when $k<1$. Instead of the black hole phase for $k>1$,
when $k<1$, the theory is described in terms of long perturbative strings,~\cite{Giveon:2005mi},
which amount,~\cite{Giveon:2019gfk,Giveon:2019twx}, in the Lorentzian theory, to the condensate
\be\label{lf}
{\cal L}_F=\lambda_F F
\ee
with $F$ being the operator in~(\ref{of})$\times (1/Q^2)^k$ and,~\cite{Giribet:2001ft,Giveon:2019gfk,Giveon:2019twx},
\be\label{lambdaf}
\lambda_F={\left(\pi\lambda\Delta(1/k)\right)^k\over\pi\Delta(k)}~,\qquad\Delta(x)\equiv{\Gamma(x)\over\Gamma(1-x)}
\ee
where $\lambda\sim{1\over g^2}$, with $g$ the value of the string coupling at the horizon,
is the graviton condensate coupling which turns the linear-dilaton cylinder with slope $Q$ to the cigar.

It was further argued in~\cite{Giveon:2019gfk,Giveon:2019twx} that the instantly-created folded strings condensate
affects the Lorentzian $SL(2)_k/U(1)$ horizon physics {\it dramatically}, even for parametrically large~$k$.
In the string phase regime, a.k.a. when $k<1$, the condensate~(\ref{lf}) dominates the $SL(2)_k/U(1)$ physics {\it completely}.

Curiously, $\lambda_F$ vanishes when $k=1/2$. A possible reasoning for this fact is the following.
For $k>1/2$, the superstring theory on the $SL(2)_k/U(1)$ SCFT is in the black-hole phase (when $k>1$) and string phase (when $1>k>1/2$),
which both have an instantly-created folded strings condensate, (\ref{lf}).
On the other hand, at $k=1/2$, the theory is in a third phase: a QM phase of the two-dimensional fermionic string
-- the fermionic quantum matrix model analog of~\cite{Kazakov:2000pm}, conjectured in~\cite{Giveon:2003wn}.
The transition from the string phase to the matrix QM phase is reflected via the $\lambda_F\to 0$ limit as $k\to 1/2$.
We may thus regard $\lambda_F(k)$ in~(\ref{lambdaf}) as an order parameter for this phase transition.

With the assumptions in~\cite{Chen:2021emg} and~\cite{Giveon:2021gsc}, motivated by~\cite{Soda:1993xc,Emparan:2013xia} and~\cite{Giveon:2020xxh},
respectively, the observations above have applications for the physics of $D$-dimensional Schwarzschild black holes with horizons size
\be\label{rzero}
r_0={D-3\over 4\pi}\beta\equiv{D-3\over 2}\sqrt{\alpha' k}
\ee
in the {\it superstring} theory. Here $\beta$ is the inverse Hawking temperature of the $D$-dimensional Schwarzschild black hole,
and $k$ is a defined parameter.

Let us first recall the picture that emerged for large black holes (a.k.a. with $k\gg 1$ in~(\ref{rzero}))
and string-scale black holes (a.k.a. with $k\sim 1$) in the bosonic string.

In~~\cite{Soda:1993xc,Emparan:2013xia}, it was shown that the large $D$ limit of a large Schwarzschild black-hole metric gives rise,
in the near-horizon region, $r-r_0\ll r_0$, to the sigma-model background of a two-dimensional $SL(2)/U(1)$ black hole,~\footnote{Upon
reduction on the sphere, $S^{D-2}$.}
with the $SL(2)$ level being the $k$ in~(\ref{rzero}) (up to corrections in $1/k$).

In~\cite{Chen:2021emg}, it was claimed that certain aspects of the near-horizon physics of large $D$ black holes in the bosonic string theory
are thus well described by the bosonic $SL(2)_{k+2}/U(1)$ CFT  (note that the bosonic $SL(2)$ level is $k+2$), for {\it any} $k$.
A non-trivial test of this claim is the match of first $\alpha'$ corrections to $\beta$, the metric and dilaton
of large $D$ black holes in the bosonic string versus those in the corresponding $SL(2)/U(1)$ sigma-model.
As a consequence, it was argued that as the Hawking temperature increases above $T={1\over 2\pi\ell_s}$, which amounts to $k=1$ in~(\ref{rzero})
(a.k.a. bosonic level $k+2=3$), the black holes are likely to turn over to highly excited strings -- strings stars.

Moreover, the exact $SL(2)_{k+2}/U(1)$ CFT was used in~\cite{Chen:2021emg} to propose
a form for the Lyapunov (or chaos) exponent $\lambda_L$ for the cigar theory.
The idea was to use the exact CFT to guess the spin of the exchanged state in a scattering experiment that happens
near the horizon, in the Lorentzian section, and translate it to $\lambda_L(k)$ for {\it any}~$k$.
In the range where the black-hole picture is applicable $(k+2>3)$,
the $\lambda_L(k)$ thus obtained was argued to take into account the full stringy correction to the Lyapunov exponent
in the large~$D$ limit. In particular, it was pointed out in~\cite{Chen:2021emg} that there is an agreement between the stringy effects in
scrambling, found using the methods of~\cite{Shenker:2014cwa}, and the large $k$ behavior of $\lambda_L(k)$ in the $SL(2)_{k+2}/U(1)$ CFT.

The calculation of $\lambda_L(k)$ in the $SL(2)_{k+2}/U(1)$ CFT is applicable also for $k<1$,
and the result was that while there is a  chaos exponent as long as $k>1/2$,
the Lyapunov exponent vanishes when the bosonic level is $5/2$: $\lambda_L(k\to 1/2)\to 0$.
While there is nothing apparently special for the bosonic string theory on $SL(2)_{k+2=5/2}/U(1)$,
it was recalled in~\cite{Chen:2021emg} that, on the other hand,
the fermionic string theory on supersymmetric $SL(2)_{k={1\over 2}}/U(1)$ is a very special case.

Above, we added the following observations to those in~\cite{Chen:2021emg}.
We presented the behaviors of the instantly-created folded strings condensate operator $F$ of~\cite{Giveon:2019gfk,Giveon:2019twx}
and the shock wave operator of~\cite{Chen:2021emg} in the suppersymmetric $SL(2)_k/U(1)$ case~\footnote{The operator $F$ and shock wave state
were presented in~\cite{Giveon:2019gfk,Giveon:2019twx} and~\cite{Chen:2021emg} in the bosonic case.}
and found that they coincide at $k=1/2$, precisely when the value of the folded strings condensate vanihes,
$\lambda_F(k=1/2)=0$.~\footnote{This is also true in the bosonic case.}
Moreover, we showed that the large $k$ behavior of $\lambda_L(k)$ is in agreement with the stringy effects
reported in~\cite{Chen:2021emg}, using the methods of~\cite{Shenker:2014cwa}, also in the fermionic string theories.

To recapitulate, we observed a coherence between properties of the supersymmetric cigar $SL(2)_k/U(1)$ CFT
and those expected for $D$-dimensional Schwarzschild black holes in the perturbative {\it superstring} theory, in various regimes of $k$,
a.k.a in various regimes of the horizons size $r_0$.

Lets elaborate on those in turn.

For Schwarzschild black holes with a stringy-scale Hawking temperature, a.k.a. with $\beta$ of order $\ell_s$ in~(\ref{rzero}),
the black holes geometries are highly misleading and the physics is expected to be that of strings stars -- a gas of strings, instead.
This is in harmony with the $SL(2)_k/U(1)$ cigar SCFT being dominated by instantly-created strings when $1/2<k<1$:
the folded strings condensate dominates the gravitons condensate in this range, as expected for a gas of perturbative strings.

At $k=1/2$, the strings condensate vanishes,
and this happens precisely at the value for which the folded strings operator coincides with the shock wave operator
and the chaos exponent vanishes,
and in which case the central charge of the $SL(2)_{1\over 2}/U(1)$ SCFT is the critical one, $c=15$ at $k=1/2$,~(\ref{c}),
and the physics is thus described by matrix QM, instead of Hagedorn physics when $k>1/2$.~\footnote{Recall~\cite{Giveon:2005mi}
that the large energy entropy when $1/2<k<1$ is $S=2\pi\ell_s E\sqrt{2-{1\over k}}$,
so while the theory has a Hagedorn spectrum in this range of $k$,
it ceases to have a Hagedorn behavior precisely at $k=1/2$.}~\footnote{We have not tried to understand what happens when $k<1/2$.}

For large black holes, a.k.a. for $k\gg 1$ ($r_0\gg D$),
the stringy effects in scrambling at the horizons, $u=0$, of black holes in $D$ spacetime dimensions,
with horizons size $r_0$ and inverse Hawking temperature $\beta$, can be obtained using the methods of~\cite{Shenker:2014cwa}.
For Schwarzschild black holes in the superstring theory,
the shock wave (or ``Pomeron''~\footnote{See the appendix.}) operator obtained is, approximately,~\footnote{See (67) in~\cite{Shenker:2014cwa}
for the case of $AdS_{D=d+1}$ black holes (recall that we set $\alpha'=2$ when not presented).}
\be\label{sw}
\Pi(q^2+\mu^2)\delta(u)e^{iq\cdot x}(q_u\partial u q_u\bar\partial u)^{1-(q^2+\mu^2)/2r_0^2}+...~,
\qquad\mu^2\equiv{2\pi(D-2)r_0\over\beta}
\ee
where the `$\dots$' stand for fermionic terms as required by worldsheet supersymmetry,
and the Pomeron propagator is~\footnote{Our
$\Pi$ is the superstring one, instead of the bosonic string propagator in~\cite{Shenker:2014cwa}; see the appendix.}
\be\label{pomeprog}
\Pi(q^2+\mu^2)=2\pi e^{i\pi(q^2+\mu^2)/2r_0^2}{\Gamma\left((q^2+\mu^2)/2r_0^2\right)\over\Gamma\left(1-(q^2+\mu^2)/2r_0^2\right)}
\ee
with $q$ being transverse momenta.~\footnote{The
sphere $S^{D-2}$ of parametrically large radius $r_0$ at the horizon
is approximated by a planar space $x_i$, $i=1,...,D-2$, and the $\delta(u)$ is an approximation.
(Although an exact behavior of the operator in $u$ and $v$ is needed for making the operator $(1,1)$,
the characteristic momenta in the missing pieces are very small compared to the $u$ momentum, $q_u$ -- the analog of $1/Q$ in~(\ref{w}).
Assuming the formal existence of a worldsheet theory on the black-hole background,
in the manipulations of~\cite{Shenker:2014cwa}, the $(1,1)$ condition was put by hand.)}

This behavior of shock waves operators amounts to the near-horizon exponential growth of amplitudes, $e^{\lambda_L t}$,
with~\footnote{See (72) in~\cite{Shenker:2014cwa} for the case of $AdS_{D=d+1}$ black holes.}
\be\label{llsw}
\lambda_L={2\pi\over\beta}\left(1-{\mu^2\over r_0^2}\right)={2\pi\over\beta}\left(1-{D-2\over D-3}{1\over k}+\dots\right)
\ee
where $k$ in the last equality is defined in terms of $r_0$ and/or $\beta$ in~(\ref{rzero}).

Consequently, the behavior of the chaos (or Lyapunov) exponent $\lambda_L$ at the horizon of large ($r_0\gg D\ell_s$) large $D$ black holes
is in harmony with the large $k$ behavior of the boost eigenvalue $b$,~(\ref{llcm})--(\ref{blargek}),
of the shock wave operator in the {\it exact} cigar worldsheet theory,~(\ref{ocm}),
also in the {\it supersymmetric} $SL(2)_k/U(1)$ CFT.~\footnote{The
${D-2\over D-3}$ discrepancy is due to order $1/D$ corrections to the near-horizon $SL(2)/U(1)$ CFT approximation at finite $D$.}~\footnote{To
compare the shock wave operator in~(\ref{sw}) to the large $k$ behavior of the chaos operator in the $SL(2)_k/U(1)$ SCFT,
note that the ``Pomeron'' operator (see the appendix) of~(\ref{wvo}) with~(\ref{vcm})--(\ref{ocm}),
after the continuation to Lorentzian space,  $z\to u$, $z^*\to v$, is
\be\label{olargek}
2\pi{\Gamma\left({1\over 2k}\right)\over\Gamma\left(1-{1\over 2k}\right)}e^{-v/\sqrt{2k}}\left(\sqrt{k\over 2}\partial u \sqrt{k\over 2}\bar\partial u\right)^{1-{1\over 2k}}
\ee
up to corrections in $1/k$, and then use~(\ref{rzero}).}

In~\cite{Giveon:2021gsc}, it was assumed that for the sake of estimating the number of explicit microstates
at the horizon of a large Schwarzschild black hole, and see what they look like,
it is sufficient to approximate the near-horizon theory of a $D$-dimensional Schwarzschild black hole with temperature $1/\beta$
by an $SL(2)_k/U(1)$ exact SCFT, with the level $k$ related to $\beta$ via~(\ref{rzero}).
The coherence between the properties of the supersymmetric cigar CFT
and those expected near Schwarzschild horizons in perturbative superstring theory,
for both string scale and large black holes,
provides more support to the assumptions of~\cite{Giveon:2021gsc},
that led to the claim that the microstates of Schwarzschild black holes in perturbative string theory amount to the modes
of long folded strings in the vicinity of their horizons.

While there isn't a bottom line,
one may be tempted to claim the following.
In perturbative string theory, a black hole of size $r_0$,~(\ref{rzero}),
is actually an object of size $r_0+$order $\ell_s$,~\footnote{Recall,~\cite{Giveon:2019gfk,Giveon:2019twx} and references therein,
that the wave function of the strings condensate has a tail outside the black hole, localized a distance of order the string length scale
near the horizon.}
a.k.a. a (maximally condensed) `strings star,' for {\it any} $r_0$:
semiclassically, it has the same properties as those of a black hole,
but stringy corrections to the horizon physics are dramatic, even when $r_0\gg\ell_s$.~\footnote{The idea is supported by the collection of facts mentioned above. E.g. recall,~(\ref{lambdaf}), that there is a non-vanishing value of the instantly-created folded strings condensate, $\lambda_F(k)\neq 0$, also in the black-hole phase (when $k>1$). And,~(\ref{llcm}), there is a non-vanishing chaos exponent, $\lambda_L(k)>0$, also in the strings phase (when $1/2<k<1$). Moreover, in the limit that the chaos exponent vanishes, $\lambda_L\to 0$ (when $k\to 1/2$), the chaos operator approaches the instantly-created folded strings one, and $\lambda_F\to 0$ as well,~(\ref{lambdal}). Namely, strings stars cease to exist ($\lambda_F(k=1/2)=0$) when the Lyapunov exponent vanishes ($\lambda_L(k=1/2)$). On top of the above, recall the following. The 2pf in the cigar CFT can be calculated~\cite{Giribet:2001ft} not only by using the graviton condensate but also by using the instantly-created folded strings condensate, $F$, instead. And, it gives rise~\cite{Giveon:2003wn} to a non-vanishing absorption cross section in the superstring on the cigar SCFT not only in the black-hole phase, but also in the strings phase. Assuming the proposed approximations for aspects of Schwarzschild SCFTs via the exact  $SL(2)_k/U(1)$ one (see between~(\ref{rzero}) and~(\ref{llsw})), it is consequently tempting to suggest that black holes in perturbative string theory are indistinguishable from strings stars.}
In a sense, {\it the surface of a black hole in string theory is not a horizon}.~\footnote{In the continuation to Euclidean spacetime,
it would mean that there is `a puncture in the cigar,' due to the winding strings condensate,
as was suggested in~\cite{Giveon:2021gsc,Brustein:2021qkj} and references therein.}

\vspace{10mm}

\section*{Acknowledgments}
I thank N.~Itzhaki for discussions.
This work was supported in part by the ISF (grant number 256/22), the BSF (grant number 2018068)
and the ISF center for excellence (grant number 2289/18).

\vspace{10mm}

\noindent
\section*{Appendix. Pomeron operator}

The Pomeron operator in flat spacetime is (3.6),(3.7) of~\cite{Brower:2006ea}:
\be\label{pomeron}
\int d^2w e^{ip_1\cdot x(w,\bar w)}e^{ip_2\cdot x(0)}\sim 2\pi e^{-i\pi\gamma}{\Gamma(-\gamma)\over\Gamma(1+\gamma)}
\left(p_1\cdot\partial x(0) p_1\cdot\bar\partial x(0)\right)^\gamma e^{i(p_1+p_2)\cdot x(0)}
\ee
where
\be\label{t}
\gamma\equiv 1+{\alpha'{\rm t}\over 4}~,\qquad {\rm t}\equiv -(p_1+p_2)^2
\ee
Exchanging this operator gives rise to the Regge behavior of scattering amplitudes.
Fot ${\rm t}=-4/\alpha'$, the residue of the pole on the r.h.s. of~(\ref{pomeron}) is the tachyon.
As $\gamma\to 1$, a.k.a. ${\rm t}\to 0$, the residue of the pole is the vertex operator of the graviton.

In the superstring, the flat space Pomeron operator is (3.22),(3.23) of~\cite{Brower:2006ea}:
\be\label{spomeron}
\int d^2wd^2\theta d^2\theta' e^{ip_1\cdot X(w,\bar w;\theta,\bar\theta)}e^{ip_2\cdot X(0;\theta'\bar\theta')}\sim\Pi(\gamma)
\left(p_1\cdot\partial x(0) p_1\cdot\bar\partial x(0)\right)^\gamma e^{i(p_1+p_2)\cdot x(0)}+\dots
\ee
where $x(w,\bar w)$ is the lowest component of the superfield $X(w,\bar w; \theta,\bar\theta)$ and the `$\dots$' stand for fermionic terms,
a.k.a. the Pomeron vertex operator has the same bosonic part as in the bosonic string,~(\ref{pomeron}),
together with fermionic terms as required by worldsheet supersymmetry,
and the Pomeron propagator is
\be\label{propagator}
\Pi(\gamma)=2\pi e^{i\pi(1-\gamma)}{\Gamma(1-\gamma)\over\Gamma(\gamma)}
\ee
so, it no longer has a tachyon pole.

The operator in equation~(\ref{sw}) is the first curvature correction to~(\ref{spomeron}),
at the horizon of a large, $r_0/D\gg\ell_s$, Schwarzschild black hole in $D$-dimensional spacetime,
for certain momenta, $p_{1,2}$.
The residue of the pole at $q^2+\mu^2=0$ in~(\ref{sw}) is the vertex operator for a shock wave mode of the graviton.

The continuation to Lorentzian space, $ix\to t$ ($z\to u$), of the operator in equation~(\ref{wvo}), for the case~(\ref{vcm})--(\ref{ocm}),
is the {\it exact} asymptotic behavior of the ``Pomeron'' (or shock wave) operator in the supersymmetric $SL(2)_k/U(1)$ black hole CFT.

\end{document}